\documentstyle[aps,preprint,epsfig]{revtex}

\begin{document}
\draft
\def\be{\begin{equation}}
\def\ee{\end{equation}}
\def\bfi{\begin{figure}}
\def\efi{\end{figure}}
\def\bea{\begin{eqnarray}}
\def\eea{\end{eqnarray}}
\title{Slow relaxation in the large $N$ model for phase ordering}

\author{Federico Corberi$^\dag$, Eugenio Lippiello$^\ddag$ 
and Marco Zannetti$^\S$}
\address{Istituto Nazionale per la Fisica della Materia,
Unit\`a di Salerno and Dipartimento di Fisica ``E.Caianiello'', 
Universit\`a di Salerno,
84081 Baronissi (Salerno), Italy}

\maketitle

\dag corberi@na.infn.it \ddag lippiello@sa.infn.it 

\S zannetti@na.infn.it

\begin{abstract}

The basic features of the slow relaxation phenomenology arising in phase
ordering processes are obtained analytically in the large $N$ model 
through the exact 
separation of the order parameter into the sum of thermal and condensation
components. The aging contribution in the response function $\chi_{ag}(t,t_w)$
is found to obey a pattern of behavior, under variation of dimensionality,
qualitatively similar to the one observed in Ising systems. There exists
a critical dimensionality $(d=4)$ above which $\chi_{ag}(t,t_w)$ is 
proportional to the defect density $\rho_D(t)$, while for $d<4$ it vanishes
more slowly than $\rho_D(t)$ and at $d=2$ does not vanish. As in the 
Ising case, this behavior can be understood in terms of the dependence on
dimensionality of the interplay between the defect density and the effective
response associated to a single defect.

\end{abstract}
 

\section{Introduction} \label{sec1}

The phase ordering processes\cite{Bray94} following the quench 
of non disorder systems
(e.g. ferromagnets) below the critical point provide a 
simplified framework for the study of slow relaxation phenomena.
In particular, aging and the off equilibrium deviation from the 
fluctuation dissipation theorem (FDT) have been studied  
numerically\cite{Barrat98,Parisi99,Corberi01} and through analytical 
treatment of solvable models\cite{Cugliandolo95,Berthier99,Lippiello00}.
The basic structure of the complex phenomenology arising
in these processes originates from the wide separation in the
time scales of fast and slow variables.
Referring to the more intuitive case of a domain forming
system, in the late stage of phase ordering the order
parameter can be assumed to be the sum of two statistically
independent components\cite{Mazenko88}
\be
\phi (\vec x,t)=\psi (\vec x,t)+\sigma (\vec x,t)
\label{i1}
\ee
the first describing equilibrium thermal fluctuation within domains
and the second off-equilibrium fluctuations due to interface motion.
From~(\ref{i1}) follows rather straightforwardly the split
of the autocorrelation function
\be
G(t,t_w)=G_{st}(t-t_w)+G_{ag}\left ( \frac{t}{t_w}\right )
\label{i2}
\ee
where $G_{st}(t-t_w)$ is the stationary time translation invariant (TTI)
contribution due to $\psi $ and $G_{ag}(t/t_w)$ the aging contribution
due to the off-equilibrium $\sigma $ degrees of freedom.
A similar structure shows up in the linear response at the time $t$
to an external random field switched on at the earlier time $t_w$
\be
\chi (t,t_w)=\chi _{st}(t-t_w)+\chi _{ag}(t,t_w).
\label{i3}
\ee
Here, the stationary contribution $\chi _{st}(t-t_w)$ satisfies the
equilibrium FDT with respect to $G_{st}(t-t_w)$, while
$\chi _{ag}(t,t_w)$ is the off-equilibrium extra response due to
the presence of the interfaces. This is 
considered\cite{Barrat98,Berthier99,Parisi99} 
to be proportional to the interface density
\be
\chi _{ag}(t,t_w)\sim \rho_I (t)
\label{i4}
\ee
where $\rho_I (t)\sim L^{-1}(t)$ and $L(t)\sim t^{1/z}$ is the typical 
domain size
with $z=2$ for non conserved order parameter\cite{Bray94}, as we shall assume
in the following. A more precise formulation of this behavior is through the
scaling relation 
\be
\chi _{ag}(t,t_w)=t_w^{-a}\hat \chi \left (\frac{t}{t_w}\right )
\label{i5}
\ee
with the exponent $a=1/2$. The implication is that $\chi _{ag}(t,t_w)$ is
negligible in the asymptotic regime $t_w\to \infty$.

Motivated by analytical results\cite{Lippiello00,Godreche00} for the one 
dimensional
Ising model which do not fit in the above scheme, recently
we have undertaken a detailed study of the behavior of the response
function under variation of space dimensionality for a system with
a scalar order parameter\cite{Corberi01}. On the 
basis of numerical simulations for 
discrete Ising spins and approximated analytical results for continuous
spins, we have arrived to a picture for the behavior of 
$\chi_{ag}(t,t_w)$ which modifies considerably
the one presented above. This is best understood by introducing
the notion of the effective response due to a single interface
and defined by
\be
\chi _{ag}(t,t_w)=\rho _I(t)\chi _{eff}(t,t_w).
\label{i6}
\ee
Then, if~(\ref{i4}) were to hold,
$\chi _{eff}(t,t_w)$ ought to be a constant.
Instead, we have found that for an Ising system this is the
case only for $d>3$, while for $d<3$ there is the power
law growth
\be
\chi _{eff}(t,t_w)\sim (t-t_w)^\alpha
\label{i7}
\ee
with numerical values for the exponent compatible with
$\alpha =(3-d)/4$. At $d=3$ the power law is replaced by logarithmic
growth, promoting $d=3$ to the role of a critical dimensionality.
This result is interesting for two reasons. The first concerns
the mechanism of the response and the role of the off-equilibrium
degrees of freedom. The power law~(\ref{i7}) reveals that for $d<3$
the aging component of the response
does not originate trivially just from the polarization
of interfacial spins, but a more complex phenomenon
producing large scale optimization of the position of domains with respect
to the  external field is at work. The second regards the overall 
behavior
of $\chi _{ag}(t,t_w)$. Putting together~(\ref{i6}) and~(\ref{i7})
the exponent $a$ in~(\ref{i5}) acquires a dependence on dimensionality
\be
a= \left \{ \begin{array}{ll}
			\frac{1}{2},   & d>3\\  
			\frac{d-1}{4}, & d<3
                   \end{array}                                        
               \right .
\label{i8}
\ee
showing that $\lim_{t_w \rightarrow \infty} \chi_{ag}(t,t_w)$ 
does not vanish as $d \rightarrow 1$. 
This indicates that $d=1$ plays the role of a lower critical dimensionality,
where the off equilibrium response becomes persistent. An interesting
consequence of this phenomenon is that
the connection between static and dynamic properties\cite{Franz98}, 
which holds for
$d>1$, is invalidated at $d=1$. It should be added that in order to
have a phase ordering process in $d=1$ thermal fluctuations within
domains must be suppressed\cite{Corberi01}.

As stated previously, the picture summarized above has been
established on the basis of a combination of exact results for the
$d=1$ Ising model, numerical results for Ising systems with $d>1$
and approximate analytical results for continuous spin systems.
It is, then, interesting to test how general is the picture.
As a step in this direction, we have considered the large $N$ model
where exact analytical calculations can be carried out\cite{Coniglio89}.
The slow relaxation properties of the large $N$ or equivalent
mean field models, arising in the quench at or below $T_C$
have been analyzed before\cite{Cugliandolo95,Godreche00bis}.
What we do here, however, goes beyond previous results since we manage 
to reproduce exactly and analytically the scenario outlined for Ising systems.
We show that in the large $N$ model one can make explicitly the
separation~(\ref{i1}) of the order parameter into the sum of two
independent components which are responsible of the stationary and
aging contributions in~(\ref{i2}). Then, we carry out analytically the
corresponding separation~(\ref{i3}) of the response function. After
introducing the notion of defect density for the large $N$ model, we
derive by analogy with~(\ref{i6}) the effective response per defect and
we find a behavior, as dimensionality is varied, qualitatively similar
to the one established for scalar systems.

The paper is organized as follows. In Section~\ref{sec2} 
the large $N$ model is defined and the main static properties are reviewed. 
In  Section~\ref{sec3} the solution of the equation of motion 
is presented and the analytical form of the autocorrelation 
function is obtained in quenches at and below the critical point.  
In Section~\ref{sec4} the splitting of the order parameter 
into independent components satisfying the requirements above 
described is carried out. In Section~\ref{sec5} the behavior 
of the integrated response function is considered in relation to 
the off equilibrium deviation from the FDT. 
Finally, concluding remarks are made in Section~\ref{sec6}.

\section{Model and static properties} \label{sec2}

We consider the purely relaxational dynamics of a system with a non
 conserved order parameter governed by the Langevin equation
\be
\frac{\partial \vec \phi (\vec x, t)}{\partial t} = 
-\frac{\delta{\cal H}[\vec \phi]}{\delta \vec \phi(\vec x,t)}
+ \vec \eta (\vec x,t)
\label{2.1}
\ee
where $\vec \phi = (\phi_1,...,\phi _N) $ is an $N$-component vector, 
$\vec \eta (\vec x,t)$ is a gaussian white noise with expectations 
\be
\left \{ \begin{array}{ll}
\langle \vec \eta (\vec x, t) \rangle = 0 \\  
\langle \eta _\alpha (\vec x,t) \eta _\beta (\vec x',t') \rangle 
= 2T \delta _{\alpha,\beta} \delta (\vec x -\vec x') \delta (t-t')
                   \end{array}                                        
               \right .
 \label{2.2}
\ee
and $T$ is the temperature of the thermal bath. 
In the dynamical process of interest the system is initially prepared 
in the infinite temperature equilibrium state with expectations  
\be
\left \{ \begin{array}{ll}
      \langle \vec \phi (\vec x) \rangle = 0  \\ 
      \langle \phi _\alpha (\vec x) \phi _\beta (\vec x') \rangle = 
      \Delta \delta _{\alpha,\beta} \delta (\vec x -\vec x') 
                       \end{array}                                        
               \right . 
\label{2.3}
\ee
and at the time $t=0$ is quenched to a lower final temperature $T_F$. 
The hamiltonian is of the Ginzburg-Landau form
\be
{\cal H} [\vec \phi] = \int _{V} d ^d x \left 
[ \frac{1}{2} ( \nabla  \vec \phi )^2 + \frac{r}{2} \vec \phi ^2 
+ \frac{g}{4N}(\vec \phi ^2)^2 \right ]
\label{2.4}
\ee 
where $r<0$, $g>0$ and $V$ is the volume of the system. 
In the large $N$ limit the equation of motion 
for the Fourier transform of the order parameter 
$ \vec \phi (\vec k) = \int _V d^d x \vec \phi(\vec x) 
\exp (i \vec k \cdot \vec x)$ takes the linear form
\be
\frac{\partial \vec \phi (\vec k ,t) }{\partial t} 
= - [k^2 + I(t) ] \vec \phi( \vec k,t) + \vec \eta (\vec k,t)
\label{2.5}
\ee
where
\be
\left \{ \begin{array}{ll}
      \langle \vec \eta (\vec k, t) \rangle = 0  \\ 
      \langle \eta _\alpha (\vec k,t) \eta _\beta (\vec k',t') \rangle = 2T_F
      \delta _{\alpha,\beta}V \delta_{\vec k +\vec k',0} \delta(t-t') 
                       \end{array}                                        
               \right . 
\label{2.6}
\ee
and the function of time 
\be
I(t) = r + \frac{g}{N} \langle \vec \phi ^2(\vec x,t)\rangle 
\label{2.7}
\ee
must be determined self-consistently, with the average on the right hand 
side taken both over thermal noise and initial condition. 
If the volume $V$ is kept finite the system equilibrates in a 
finite time $t_{eq}$ and the order parameter probability distribution
 reaches the Gibbs state 
\be
P_{eq} [\vec \phi(\vec k)] = \frac {1}{Z} e^{ -\frac {1}{2 T_F V} 
\sum _{\vec k} ( k^2+ \xi ^{-2}) \vec \phi(\vec k) \cdot \vec \phi(-\vec k)  }
\label{2.8}
\ee
where $ \xi $ is the correlation length defined by 
the equilibrium value of $I(t) $ through 
\be
\xi ^{-2} = r+\frac{g}{N} \langle \vec \phi ^2(\vec x) \rangle _{eq}
\label{2.9}
\ee
with $\langle \cdot \rangle _{eq}$ standing for the average 
taken with~(\ref{2.8}). 

In order to analyze the properties of
$P_{eq} [\vec \phi(\vec k)]$ it is necessary to extract from~(\ref{2.9})
the dependence of $\xi ^{-2}$ on $T$ and $V$. Evaluating the average,
the above equation yields 
\be
\xi ^{-2} = r+\frac{g}{V} \sum _{\vec k} \frac{T_F}{k^2+\xi ^{-2}}. 
\label{2.10}
\ee
The solution of this equation is well known\cite{Baxter} and here we summarize 
the main features. Separating the $\vec k=0$ term under 
the sum, for very large volume we may rewrite
\be
\xi ^{-2} = r+g T_F B(\xi ^{-2}) + g \frac{T_F}{V \xi ^{-2}}
\label{2.11}
\ee
 where 
\be
B(\xi ^{-2}) = \lim _{V \to \infty} \frac{1}{V} 
\sum _{\vec k} \frac {1}{k^2+ \xi ^{-2}}  
= \int \frac{d^d k}{(2 \pi)^d} \frac{e^{-\frac{k^2}{\Lambda^2}}}{k^2+\xi^{-2}}
\label{2.12}
\ee
regularizing the integral by introducing the high momentum
cutoff $\Lambda $.
The function $B(x)$ is a non 
negative monotonically decreasing function with the maximum value at $x=0$ 
\be
B(0)=\int \frac{d^d k}{(2 \pi)^d} \frac{e^{-\frac{k^2}{\Lambda^2}}}{k^2}
= (4 \pi)^{-\frac{d}{2}} \frac{2}{d-2} \Lambda^{d-2} \quad .
\label{2.13}
\ee
By graphical analysis one can easily show that~(\ref{2.11}) 
admits a finite solution for all $T_F$. However, there exists 
the critical value of the temperature $T_C$ defined by 
\be
r+gT_C B(0) =0
\label{2.14}
\ee
such that for $T_F>T_C$ the solution is independent of the volume,
while for $T_F \leq T_C$ it depends on the volume. Using
\be
B(x) =(4 \pi)^{-\frac{d}{2}}
x^{\frac{d}{2}-1}e^{\frac{x}{\Lambda^2}}
\Gamma \left ( 1-\frac{d}{2},\frac{x}{\Lambda^2} \right )
\label{2.14.1}
\ee
where $\Gamma(1-\frac{d}{2},\frac{x}{\Lambda^2})$ is the incomplete 
gamma function, for $0 < \frac{T_F-T_C}{T_C} \ll 1$
one finds (Appendix I) $\xi \sim \large ( \frac{T_F-T_C}{T_C}
\large )^{-\nu}$ where $\nu = 1/2$ for $d>4$ and 
$\nu = 1/(d-2)$ for $d<4$, with logarithmic corrections for
$d=4$. At $T_C$ one has $\xi \sim V^{\lambda}$ with $\lambda = 1/4$
for $d>4$ and $\lambda = 1/d$ for $d<4$, again with logarithmic
corrections for $d=4$. Finally, below $T_C$ one finds $\xi^2 =
\frac {M^2 V}{T_F}$ where
$M^2= M_0^2 \left ( \frac {T_C - T_F}{T_C} \right )$
and $M_0^2 = -r/g$.

Let us now see what are the implications for the equilibrium state. 
As~(\ref{2.8}) shows the individual Fourier components are 
independent random variables, gaussianly distributed with zero 
average. The variance is given by 
\be
\frac{1}{N} \langle \vec \phi(\vec k) \cdot \vec \phi(-\vec k) \rangle _{eq} = 
V C_{eq}(\vec k)
\label{2.18}
\ee
where 
\be
C_{eq}(\vec k) = \frac{ T_F}{k^2+ \xi^{-2}}
\label{2.19}
\ee
is the equilibrium structure factor. For $T_F >T_C$, 
all $\vec k$ modes behave in the same way, 
with the variance growing linearly with the volume. For $T_F \leq T_C$, 
instead, $\xi^{-2}$ is negligible
with respect to $k^2$ except at $\vec k=0$, yielding  
\bea
C_{eq}(\vec k)  =   \left \{   
                       \begin{array}{ll}
                          \frac{ T_C}{k^2} (1 -\delta _{\vec k,0}) + 
                          a V^{2\lambda} \delta _{\vec k,0}      & 
                                 \mbox{, for $T_F=T_c$ }\\ 
        \frac{T_F}{k^2 }(1-\delta _{\vec k,0})+ M^2 V \delta _{\vec k,0} &
				 \mbox{, for $T_F<T_c$} 
                       \end{array}
		    \right .
\label{2.20}
\eea
where $a$ is a constant (Appendix I). This produces
a volume dependence in the variance of the $\vec k=0$ mode
growing faster than linear. Therefore, for $T_F \leq T_C$ the
$\vec k=0$ mode behaves differently from all the other modes with $\vec k \neq 0$.
For $T_F<T_C$ the probability distribution~(\ref{2.8}) 
takes the form 
\be
P_{eq}[\vec \phi (\vec k) ] = \frac {1} {Z} 
e ^{-\frac{\vec \phi ^2(0)}{2 M^2 V^2}}  
e^{ -\frac{1}{2T_FV} \sum_{\vec k} k^2 \vec \phi (\vec k) \cdot
\vec \phi (-\vec k) }
\quad .
\label{2.21}
\ee
Therefore, crossing $T_C$ there is a transition from the usual disordered
high temperature phase to a low temperature phase characterized by a 
macroscopic variance in the distribution of the $\vec k=0$ mode.
The distinction between this phase and the mixture of pure states,
obtained below $T_C$ when $N$ is kept finite, has been discussed  
elsewhere\cite{Castellano97}.

We shall refer to this transition as condensation of fluctuations
in the $\vec k=0$ mode. In order to gain a better insight it is
convenient to go back to real space, splitting the order parameter
into the sum of the two independent components
\be
\vec \phi (\vec x) = \vec \sigma + \vec \psi (\vec x) 
\label{2.22}
\ee
with
\be
\vec \sigma = \frac{1}{V}\vec \phi (\vec k=0)
\label{2.23}
\ee
and 
\be 
\vec \psi (\vec x) = \frac{1}{V} \sum _{\vec k \neq 0 } 
\vec \phi (\vec k) e^{i\vec k \cdot \vec x}.
\label{2.24}
\ee
Then~(\ref{2.21}) takes the form 
\be
P_{eq}[\vec \phi (\vec x) ] = P(\vec \sigma) P[\vec \psi (\vec x) ] 
\label{2.24.1}
\ee
where   
\be
P(\vec \sigma) = \frac{1}{(2 \pi M^2)^{N/2}} e^{- \frac{\vec \sigma ^2  }
{2 M^2}} \quad 
\label{2.25}
\ee
shows the formation of the condensate below $T_C$ with the macroscopic
variance $\frac{1}{N} \langle \vec \sigma^2 \rangle _{eq} = M^2$
while
\be
P[\vec \psi (\vec x)] = \frac{1}{Z} e^{-\frac{1}{2T_F}
\int_V d^d x (\nabla \vec \psi )^2}
\label{2.25.1}
\ee
describes thermal fluctuations about the condensate. Correspondingly,
the correlation function 
$ G_{eq}(\vec x-\vec x') = (1/N) \langle \vec \phi (\vec x) \cdot 
\vec \phi (\vec x') \rangle_{eq} $ splits into the sum of two pieces
\be
G_{eq}(\vec x- \vec x') =  G_T(\vec x- \vec x') + M^2
\label{2.26}
\ee
where 
\be
G_{T}(\vec x- \vec x') =  \frac{1}{N} 
\langle \vec \psi (\vec x) \cdot \vec \psi (\vec x') \rangle _{eq} 
\label{2.27}
\ee
is the correlation of thermal fluctuations and $M^2$ comes from
fluctuations of the condensate. Since $G_{T}(\vec x- \vec x')$
at large distances decays like $|\vec x- \vec x' |^{2-d}$,
from~(\ref{2.26}) follows $\lim_{\mid \vec x- \vec x' \mid 
\rightarrow \infty} G_{eq}(\vec x- \vec x') = M^2$ showing the
violation of the clustering property of the correlation function
due to the breaking of ergodicity in the low temperature phase.
For future reference, notice that from~(\ref{2.9}) follows that
for $\vec x = \vec x'$ and $T_F \leq T_C$
\be
G_{eq}(0) = M_0^2
\label{2.27.1}
\ee
which in turn implies
\be
G_{T}(0) = M_0^2-M^2.
\label{2.27.2}
\ee

As stated above, with a finite $V$ equilibrium is reached
for $t \sim t_{eq}$ and $t_{eq} \sim \xi^2$. Hence, in a quench to 
$T_F < T_C$ one has $t_{eq} \sim V$ implying that if the $V \rightarrow
\infty$ limit is taken at the beginning of the quench, equilibrium is
not reached for any finite time. Since in the following we are interested
in the relaxation regime before equilibrium is reached, we shall take the
thermodynamic limit from the outset. We are interested
to see whether it is possible to carry out the decomposition~(\ref{i1})
of the order parameter in such a way that $\vec \sigma (\vec x,t)$
eventually evolves into the equilibrium condensate $\vec \sigma$ and
$\vec \psi (\vec x,t)$ into the equilibrium thermal fluctuations.

\section{Dynamics} \label{sec3}

Due to rotational symmetry, from now on we shall drop vectors
and refer to the generic component of the order parameter.
The formal solution of~(\ref{2.5}) is given by
\be
\phi (\vec k,t) = R(\vec k,t,0)\phi _0(\vec k)+
\int _0 ^t dt' R(\vec k,t,t')\eta (\vec k,t')
\label{3.5}
\ee
where $\phi _0(\vec k)=\phi (\vec k,t=0)$ and according to~(\ref{2.3})
in the $V \rightarrow \infty$ limit
\be
\left \{ \begin{array}{ll}
      \langle \phi _0 (\vec k)\rangle  = 0  \\ 
       \langle \phi _0(\vec k)\phi _0(\vec k')\rangle  = \Delta (2\pi)^d
                                                   \delta(\vec k+\vec k'). 
                       \end{array}                                        
               \right . 
\label{3.4}
\ee
The response function is given by
\be
R(\vec k,t,t')=\frac{Y(t')}{Y(t)}e^{-k^2(t-t')}
\label{3.6}
\ee
with $Y(t)=\exp [Q(t)]$, $Q(t)=\int _0 ^t ds I(s)$
and $Y(0)=1$. 
The actual solution is obtained once the function $Y(t)$ is determined.
In order to do this, notice that from the definition of $Y(t)$ follows
\be
\frac{dY^2(t)}{dt}=2\left [ r+g\langle \phi ^2(\vec x,t)\rangle \right ]
Y^2(t).
\label{3.9}
\ee
Writing $\langle \phi ^2(\vec x,t)\rangle $ in terms of the 
structure factor
\be
\langle \phi ^2(\vec x,t)\rangle =\int \frac {d^dk}{(2\pi )^d}
C(\vec k,t)e^{-\frac{k^2}{\Lambda ^2}}
\label{3.10}
\ee
and using~(\ref{3.5}) to evaluate $C(\vec k,t)$  
\be
C(\vec k,t)=R^2(k,t,0)\Delta+2T_F\int _0 ^t dt' R^2 (\vec k,t,t')
\label{3.11}
\ee
from~(\ref{3.9}) we obtain the integro-differential 
equation 
\be
\frac{dY^2(t)}{dt}=2rY^2(t)+2g\Delta f\left (t+\frac{1}{2\Lambda ^2}\right )
+4gT_F\int _0 ^t dt' f\left (t-t'+\frac{1}{2\Lambda ^2}\right ) Y^2(t')
\label{3.12}
\ee
where
\be
f(x)\equiv \int \frac{d^dk}{(2\pi )^d}e^{-2k^2x}=(8\pi x)^{-\frac{d}{2}}.
\label{3.13}
\ee
Solving~(\ref{3.12}) by Laplace transform\cite{Newman90,Godreche00bis}, 
the leading behavior of $Y(t)$
for large time is given by
\be
Y^2(t)=\left \{ \begin{array}{ll}
                              A_ae^{2 \xi^{-2}t}    & \mbox{; for $T_F>T_C$} \\
                              A_ct^{\omega} & \mbox{; for $T_F=T_C$} \\
			      A_bt^{-\frac{d}{2}}  & \mbox{; for $T_F<T_C$}
                              \end{array}
       \right .
\label{3.14}
\ee
where $\omega = 0$ for $d>4$ and $\omega =(d-4)/2$ for $d<4$. The values of the
constants $A_a, A_b$ and $A_c$ are listed in Appendix II.

This completes the solution of the model. Once the response function is known,
we may go back to~(\ref{3.5}) and take various averages. Using~(\ref{2.6})
and~(\ref{3.4}) we have
\be
\langle \phi (\vec k,t)\rangle =0  
\label{8.1}
\ee
and for the two time structure factor $\langle \phi (\vec k,t)
\phi (\vec k',t') \rangle = C(\vec k,t,t')(2\pi )^d
\delta (\vec k+\vec k')$ 
with $t\geq t'$
\be
C(\vec k,t,t')=R(\vec k,t,0)R(\vec k,t',0)\Delta+
2T_F\int _0 ^{t'} dt'' R(\vec k,t,t'')R(\vec k,t',t'').
\label{8.2}
\ee
The corresponding real space correlation function
$G(\vec x-\vec x',t,t')=\langle \phi (\vec x,t)
\phi (\vec x',t') \rangle$ 
is given by
\bea
G(\vec x-\vec x',t,t')     &=& \int d \vec x'' R(\vec x-\vec x'',t,0)
                                  R(\vec x'-\vec x'',t',0)\Delta \nonumber \\
                           &+& 2T_F\int _0 ^{t'} dt'' \int d \vec x''
                           R(\vec x-\vec x'',t,t'')R(\vec x'-\vec x'',t',t'')  
\label{9.1}
\eea
where $R(\vec x,t,t')$ is the inverse Fourier transform of $R(\vec k,t,t')$.
In the following we will be primarily concerned with the autocorrelation
function $G(t,t')=G(\vec x-\vec x'=0,t,t')$. Using the 
definitions~(\ref{3.6}) and~(\ref{3.13}) this is given by
\be
G(t,t')=\frac{1}{Y(t)Y(t')}\left [ f\left (\frac{t+t'}{2}+\frac{1}{2\Lambda ^2}
\right )\Delta +2T_F\int _0 ^{t'} dt'' f\left (\frac{t+t'}{2}-t''+
\frac{1}{2\Lambda ^2}\right ) Y^2(t'') \right ].
\label{9.2}
\ee
The behavior of this quantity for different final temperatures
and for different time regimes has been studied in 
the literature\cite{Newman90,Godreche00bis}. 
Here we summarize the results.

For $T_F > T_C$ 
from~(\ref{3.14}) follows that for $t' > t_{eq}=2 \xi^{-2}$
the autocorrelation function is TTI
\be
G(\tau)=G_{eq}(0)e^{-\frac {\tau}{t_{eq}}}
\label{11.1}
\ee
where $G_{eq}(0)$ is the equilibrium fluctuation above $T_C$ 
given by~(\ref{2.9}), i.e. $G_{eq}(0)= M_0^2 +\frac{1}{g}\xi^{-2}$
and $\tau=t-t'$. 

For $T_F\leq T_c$ the equilibration time diverges and there are
two time regimes of interest:

{\it i}) short time separation:  
$t'\to \infty \quad , \quad \frac{\tau } {t'} \to 0$

{\it ii}) large time separation:  
$t'\to \infty \quad , \quad \frac{\tau } {t'} \to \infty$.

\begin{figure}
\centerline{\psfig{figure=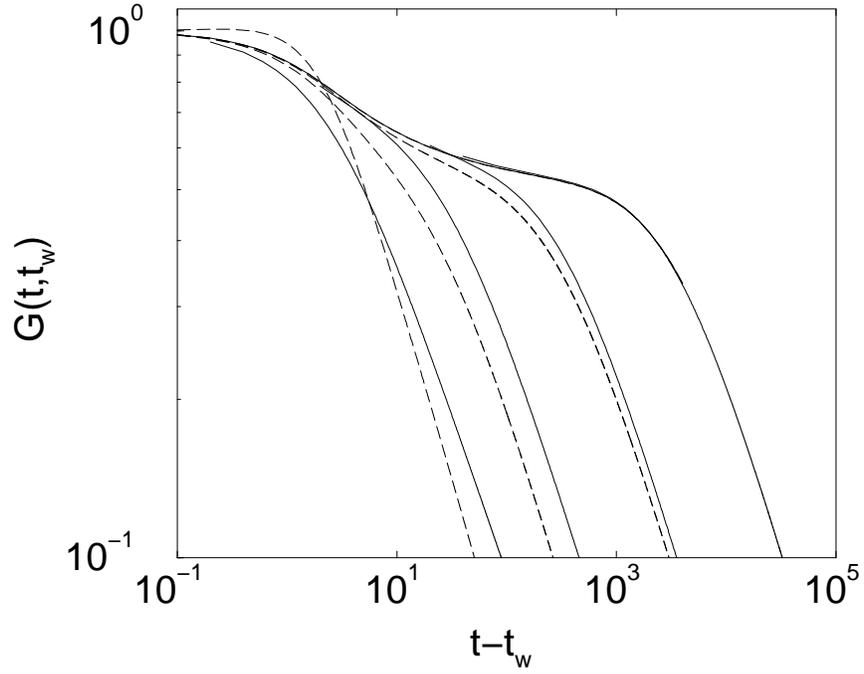,width=9cm,angle=-90}}
\caption{Comparison of the exact autocorrelation function $G(t,t_w)$
(broken line) with the sum $G_{st}(t-t_w)+ G_{ag}(t/t_w)$
(continous line) for $t_w=1,10,10^2,10^3$ increasing from left to right.
For $t_w=10^3$ the two curves are indistinguishable.
Parameters of the quench are $d=3, T_F=T_C/2$ and $\Delta=1$.}
\label{autocorr1}
\end{figure}

\noindent Taking $t'$ large and using~(\ref{3.14}), for $T_F=T_C$
in these limits we find 
\be
G(t,t') =\left \{ \begin{array}{ll}
                               M_0 ^2 (\Lambda ^2 \tau+1)^{1-\frac{d}{2}} 
                               & \mbox{, for $\frac{\tau}{t'}\to 0$} \\
                               At'^{1-\frac{d}{2}}F(\frac{t}{t'})         
                               & \mbox{, for $\frac{\tau }{t'}\to \infty$} 
                 \end{array}
       \right .
\label{53}
\ee
with
\be
F(x)=  \left \{ \begin{array}{ll}
       \frac{4}{(4\pi)^{d/2}(d-2)}(x-1)^{1-d/2}\frac{x^{1-d/4}}{x+1}
					& \mbox{, for $2<d<4$ } \\ 
       \frac{4}{(8\pi)^{d/2}(d-2)}\left [ \left ( \frac{x-1}{2}\right )^{1-d/2}
       -\left ( \frac{x+1}{2}\right )^{1-d/2}\right ]		
					& \mbox{, for $d>4$ } \\ 
                 \end{array}                                        
               \right . 
\label{54}
\ee
and for $T_F<T_c$
\be
G(t,t') =\left \{ \begin{array}{ll}
                      M^2 +(M_0^2-M^2)(\Lambda ^2\tau+1)^{1-\frac{d}{2}}
                      & \mbox{, for $\frac{\tau}{t'}\to 0$} \\
                      M^2\left [ \frac{4tt'}{(t+t')^2}\right ]^{\frac{d}{4}}
                      & \mbox{, for $\frac{\tau }{t'}\to \infty$}.
                 \end{array}
       \right .
\label{55}
\ee
In the latter case we may also write
\be
G(t,t')=G_{st}(\tau)+G_{ag}\left (\frac {t'}{t}\right )
\label{56}
\ee
where
\be
G_{st}(\tau)=(M_0 ^2-M^2) (\Lambda ^2 \tau+1)^{1-\frac{d}{2}}
\label{57}
\ee
and
\be
G_{ag}(x)=M^2\left [ \frac{4x}{(1+x)^2}\right ]^{\frac{d}{4}}.
\label{58}
\ee
This is illustrated in Fig.\ref{autocorr1} which shows the convergence
toward the form~(\ref{56}) of the exact autocorrelation function
$G(t,t')$ obtained by solving numerically the coupled set of
equations~(\ref{3.12}) and~(\ref{9.2}). 

As explained in the Introduction, this behavior is suggestive of the existence 
of two variables responsible respectively of the stationary and of the 
aging behaviors. We want now to show that in the large $N$ limit 
these two variables can be explicitly constructed.

\section{Splitting of the field} \label{sec4}

The task stated at the end of the previous Section
requires the splitting of 
the order parameter field into the sum
of two independent contributions
\be
\phi (\vec x,t) =\psi (\vec x,t) +\sigma(\vec x,t)
\label{4.1}
\ee
with zero averages $\langle \psi (\vec x,t)\rangle = 
\overline {\sigma (\vec x,t)}=0$ and autocorrelation functions
such that 
\be
\langle \psi (\vec x,t)\psi (\vec x,t')\rangle =G_{st}(\tau )
\label{4.2}
\ee
\be
\overline {\sigma (\vec x,t)\sigma (\vec x,t')}=G_{ag}(t'/t).
\label{4.3}
\ee
In order to stress the statistical independence of the
two component fields, we have used the angular brackets for averages over
$\psi $ and the overbar for averages over $\sigma $.

For the construction of fields 
with these properties, let us go back to Fourier space. Using the
multiplicative property of the response function
\be
R(\vec k,t,t') R(\vec k,t',t_0)= R(\vec k,t,t_0)
\label{4.4}
\ee
with $t>t'>t_0$ it is easy to show that the formal solution~(\ref{3.5})
of the equation of motion can be rewritten as the sum of two statistically
independent components
$\phi (\vec k,t) =\psi (\vec k,t)+\sigma(\vec k,t)$
with
\be
\sigma(\vec k,t)= R(\vec k,t,t_0)\phi (\vec k,t_0)
\label{4.6}
\ee
and
\be
\psi (\vec k,t)=\int _{t_0}^t dt' R(\vec k,t,t')\eta (\vec k,t')
\label{4.7}
\ee
since for $0\leq t_0<t$, $\phi (\vec k,t_0)$ and $\eta (\vec k,t)$ are
independent by causality. In other words, the order parameter at the
time $t$ is split into the sum of a component $\sigma (\vec k,t)$
driven by the fluctuations of the order parameter at the earlier time
$t_0$ and a component $\psi (\vec k,t)$ driven by the thermal history
between $t_0$ and $t$.
Let us remark that $t_0$ can be chosen arbitrarily between the initial 
time of the quench ($t=0$) and the observation time $t$. With the
particular choice $t_0=0$, the component $\sigma (\vec k,t)$ is
driven by the fluctuations in the initial condition~(\ref{3.4}). 
The $\psi $ component
describes fluctuations of thermal origin while the $\sigma $ component,
as it will be clear below, if $t_0$ is chosen sufficiently large, 
describes the local
condensation of the order parameter.

According to definitions~(\ref{4.6}) and~(\ref{4.7}), 
from~(\ref{2.6}) and~(\ref{3.4}) follows
$\overline {\sigma (\vec k,t)}=\langle \psi (\vec k,t)\rangle=0$.
The two time structure factor splits into the sum
\be
C(\vec k,t,t')=C_\sigma (\vec k,t,t')+C_\psi (\vec k,t,t')
\label{4.11}
\ee
with $\overline {\sigma (\vec k,t)\sigma (\vec k',t') }=
C_\sigma (\vec k,t,t')(2\pi)^d\delta (\vec k+\vec k')$
and
\be
C_\sigma (\vec k,t,t')=
R(\vec k,t,t_0) R(\vec k,t',t_0)C(\vec k,t_0)
\label{4.12}
\ee
where $C(\vec k,t_0)$ is the equal time structure factor at the time $t_0$.
Similarly $\langle \psi (\vec k,t)\psi (\vec k',t')\rangle=
C_\psi (\vec k,t,t')(2\pi)^d\delta (\vec k+\vec k')$
with
\be
C_\psi (\vec k,t,t')
=2T_F\int _{t_0}^{t'}dt'' R(\vec k,t,t'')R(\vec k,t',t'').   
\label{4.13}
\ee
Going to real space and setting $\vec x=\vec x'$, we have
\be
G(t,t')=G_\sigma (t,t')+G_\psi (t,t')
\label{4.14}
\ee
with 
\be
G_\sigma (t,t')=\frac{Y^2(t_0)}{Y(t)Y(t')}\int \frac{d^dk}{(2\pi )^d}
e^{-k^2\left (t+t'-2t_0+\frac{1}{\Lambda ^2}\right ) }C(\vec k,t_0)
\label{4.15}
\ee
and
\be
G_\psi (t,t')=2T_F\int _{t_0}^{t'} dt''\frac{Y^2(t'')}{Y(t)Y(t')}
\int \frac{d^dk}{(2\pi )^d}e^{-k^2\left (t+t'-2t''+\frac{1}{\Lambda ^2}
\right ) }.
\label{4.16}
\ee
Assuming that $t$ and $t'$ are sufficiently larger than $t_0$ so that 
$C(\vec k,t_0)$ under the integral in~(\ref{4.15}) can be replaced by its 
value at $\vec k=0$ and using~(\ref{3.13}) we may write
\be
G_\sigma (t,t')=\frac{Y^2(t_0)}{Y(t)Y(t')}f\left ( \frac{t+t'}{2}-t_0
+\frac{1}{2\Lambda ^2}\right ) C_0
\label{4.17}
\ee
\be
G_\psi (t,t')=\frac{2T_F}{Y(t)Y(t')}\int _{t_0}^{t'} dt''
f\left ( \frac{t+t'}{2}-t''
+\frac{1}{2\Lambda ^2}\right ) Y^2(t'')
\label{4.18}
\ee
where $C_0=C(\vec k=0,t_0)$.

Let us now evaluate these results for short and large time separations.
In the first case the dominant contributions are given by
\be
G_\sigma (t,t')=M^2-(M_0^2-M^2)(2\Lambda ^2t_0)^{1-\frac{d}{2}}
\label{4.26}
\ee
and
\be
G_\psi (t,t')=(M_0^2-M^2)\left [(2\Lambda ^2t_0)^{1-\frac{d}{2}}
+(\Lambda ^2\tau+1)^{1-\frac{d}{2}} \right ]
\label{4.27}
\ee
where the unknown constant $C_0$ entering~(\ref{4.17}) has been
eliminated imposing that the equilibrium sum rule 
$G_{\sigma}(0)+G_{\psi}(0) = M_0^2$ be satisfied.
The sum of the above contributions is independent of $t_0$, 
as it should, and coincides
with~(\ref{55}). Similarly, in the large time regime 
we find the $t_0$ dependent results
\be
G_\sigma (t,t')=\left [M^2-(M_0^2-M^2)
(2\Lambda ^2t_0)^{1-\frac{d}{2}} \right ]
\left [ \frac{4tt'}{(t+t')^2}\right ] ^{\frac{d}{4}}
\label{4.28}
\ee
and
\be
G_\psi (t,t')=(M_0^2-M^2)(2\Lambda ^2t_0)^{1-\frac{d}{2}}
\left [ \frac{4tt'}{(t+t')^2}\right ] ^{\frac{d}{4}}
\label{4.29}
\ee
again with the sum independent of $t_0$ and giving back~(\ref{55}).
Defining the microscopic time $t^*=\Lambda ^{-2}$, we see that taking
$t_0\gg t^*$ from~(\ref{4.26}) and~(\ref{4.27}) we get the short time behavior
\be
\left \{ \begin{array}{ll}
	G_\sigma (t,t')=M^2 \\
	G_\psi (t,t') = (M_0^2-M^2)(\Lambda ^2\tau+1)^{1-\frac{d}{2}}
          \end{array}
       \right .
\label{4.31}
\ee
while from~(\ref{4.28}) and~(\ref{4.29}) follows the large time behavior
\be
\left \{ \begin{array}{ll}
G_\sigma (t,t')= M^2 \left [ \frac{4tt'}{(t+t')^2}\right ] ^{\frac{d}{4}} \\
G_\psi (t,t')= 0.
          \end{array}
       \right .
\label{4.32}
\ee
Comparing with~(\ref{57}) and~(\ref{58}) we can make the identifications
\be
G_\psi (t,t')=G_{st}(\tau )
\label{4.34}
\ee
and
\be
G_\sigma (t,t')=G_{ag} \left (\frac{t'}{t} \right ).
\label{4.35}
\ee
Therefore, the fields $\sigma(\vec x,t) $ and $\psi(\vec x,t) $ 
defined by~(\ref{4.6})
and~(\ref{4.7}), with the choice of $t_0 \gg t^*$,
provide an explicit realization of the decomposition~(\ref{4.1})
satisfying the requirements~(\ref{4.2}) and~(\ref{4.3}).
The physical meaning of the two components can be readily understood
comparing the equal time values of~(\ref{4.31}),
which yield respectively $G_\sigma (t=t')=M^2$ and
$G_\psi (t=t')=M_{0}^2-M^2$, with the equilibrium
results~(\ref{2.27.1}) and~(\ref{2.27.2}). It is then clear that the field 
$\sigma(\vec x,t) $ is associated to 
local condensation of the order parameter with fluctuations of size $M^2$, 
while
the field $\psi(\vec x,t) $ describes thermal fluctuations.
In the case of a system with a scalar order parameter, $\sigma $ 
would be associated to the average value of the
order parameter characteristic of a domain and $\psi $ to thermal 
fluctuations within domains~\cite{Corberi01}. 
This makes also clear the origin 
of the requirement $t_0 \gg t^*$. In order to make a separation
of variables with the above physical meaning, it is necessary
to wait a time $t_0$ large enough for all microscopic 
transients to have occurred leaving
well formed local equilibrium.

\section{Response function} \label{sec5}

In the previous Section we have produced the explicit separation
of the order parameter into the condensation component and thermal
fluctuations in the quench at $T_F< T_C$. It is
now interesting to see in what relation these components
are with the linear response function. 

\begin{figure}
\centerline{\psfig{figure=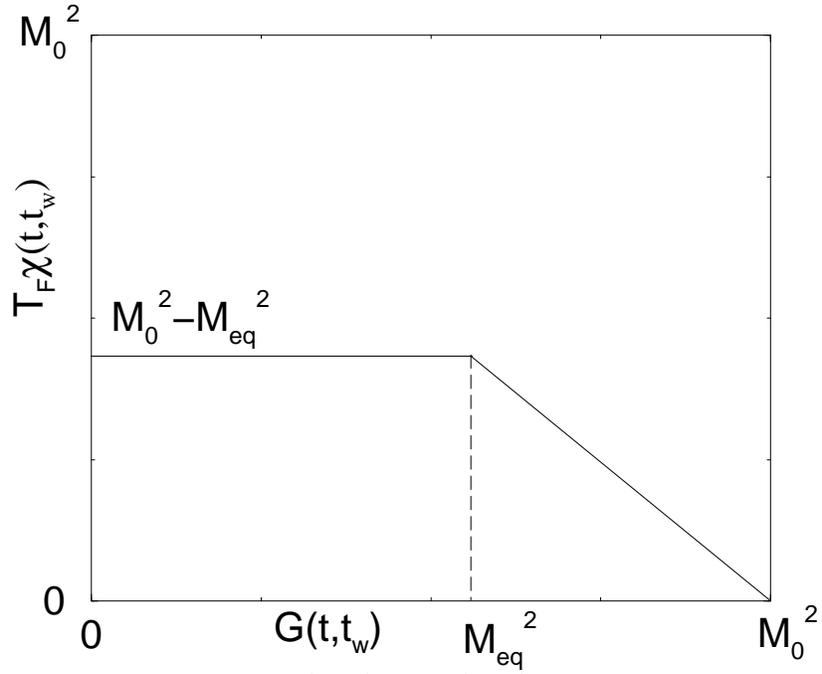,width=9cm,angle=-90}}
\caption{Parametric plot of $T_F\chi(t,t_w)$ vs. $G(t,t_w)$ in
the limit $t_w \rightarrow \infty$ for $d>2$.}
\label{phas_ord_fdt}
\end{figure}

If an external field is switched on at the time $t_w>t_0$, the 
splitting~(\ref{4.1}) modifies into $\vec \phi_h (\vec x,t)=
\vec \psi (\vec x,t) + \vec \sigma _h (\vec x,t)$ where to linear
order
\be
\vec \sigma _h (\vec x,t)=\vec \sigma (\vec x,t)+\int _{t_w}^t
dt' \int _V d\vec x' R(\vec x-\vec x',t,t')\vec h(\vec x')
\label{ottantasei}
\ee
with $\sigma (\vec x,t), \psi (\vec x,t)$ and $R(\vec x,t,t')$ unperturbed 
quantities. Here, we are interested in 
the response to a quenched, gaussianly distributed random field
with expectations
\bea
\left \{ \begin{array}{ll}
E_h[\vec h (\vec x)]=0 \\ 
E_h[h _\alpha (\vec x) h _\beta (\vec x') ]
= h_0^2 \delta_{\alpha \beta}\delta (\vec x -\vec x'). 
          \end{array}
       \right .
\label{5.2}
\eea
Computing the staggered magnetization from~(\ref{ottantasei}) and averaging
over the field we obtain
\be
\frac{1}{Nh_0^2 V}\int_V d \vec x E_h[\overline {\vec \sigma_h (\vec x,t)}
\cdot \vec h(\vec x)] = \chi(t,t_w)
\label{5.3}
\ee
where $\chi(t,t_w)= \int_{t_w}^t R(t,t')$ is the integrated response
function and $R(t,t')=R(\vec x - \vec x'=0,t,t')$.
From~(\ref{3.6}) and~(\ref{3.14}) for $T_F<T_c$
\be
R(t,t')=\int \frac {d^dk}{(2\pi )^d}R(\vec k,t,t')e^{-\frac{k^2}{\Lambda ^2}}
=(4\pi )^{-\frac{d}{2}} \left (\frac{t'}{t}\right )^{-\frac{d}{4}}
\left (t-t'+\frac{1}{\Lambda ^2}\right )^{-\frac{d}{2}}.
\label{5.1b}
\ee
Let us then write the integrated response function 
as the sum
\be
\chi (t,t_w)=\chi_{st} (t-t_w) +\chi_{ag} (t,t_w)
\label{5.4}
\ee
where the stationary component $\chi _{st}(t-t_w)$ 
is defined by requiring that the equilibrium FDT be satisfied with 
respect to the stationary component~(\ref{4.34}) of the autocorrelation 
function, namely
\be
T_F \chi_{st} (t-t_w) = G_\psi(0) - G_\psi(t-t_w).
\label{5.5}
\ee
It is straightforward to check that this is verified by 
\bea
\chi _{st}(t-t_w)&=&(4\pi)^{-\frac{d}{2}}\int _{t_w}^t dt' 
                    \left (t-t'+\frac{1}{\Lambda ^2}\right )^{-\frac{d}{2}}
								\nonumber \\ 
                 &=&\frac{1}{T_F}(M_0^2-M^2)\left \{ 1-[\Lambda ^2
                    (t-t_w)+1]^{1-\frac{d}{2}}\right \}. 
\label{5.4b}
\eea
The aging component then remains defined by the difference
\bea
\chi _{ag}(t,t_w)&=&\chi (t,t_w)-\chi _{st}(t-t_w) \nonumber \\
                 &=&(4\pi)^{-\frac{d}{2}}t_w^{1-\frac{d}{2}}x^{\frac{d}{2}-1}
\int _x ^1  dy \left ( 1-y+\frac{x}{\Lambda ^2 t_w}\right)^{-\frac{d}{2}}
\left ( y^{-\frac{d}{4}}-1\right ) 
\label{5.6}
\eea
where $x=t_w/t$.
This shows that for $d>2$ and any fixed value of $x$ 
$\lim _{t_w\to\infty }\chi _{ag}(t,t_w)=0$ implying 
\be
\lim _{t_w\to\infty }\chi (t,t_w)=\chi _{st}(t-t_w).
\label{5.7}
\ee
Hence, in the limit $t_w\to \infty $
the equilibrium FDT~(\ref{5.5}) is satisfied
by the whole response function and
the plot of $\chi (t,t_w)$ vs $G_\psi (t,t_w)$ is linear.
Instead, if $\chi (t,t_w)$ is plotted against the full autocorrelation 
function~(\ref{4.14}), from~(\ref{56}) follows
\be
\lim _{t_w\to \infty }T_F \chi (t,t_w)=\left \{ \begin{array}{ll}
            G(t,t)-G(t,t_w)  & \mbox{, for $M^2<G(t,t_w)\leq M_0^2$} \\
            M_0^2-M^2        & \mbox{, for $G(t,t_w)<M^2$} 
            \end{array}
       \right .
\label{5.10}
\ee
yielding the behavior of Fig.\ref{phas_ord_fdt} which is  
characteristic of the phase ordering process\cite{Barrat98,Parisi99}.

\begin{figure}
\centerline{\psfig{figure=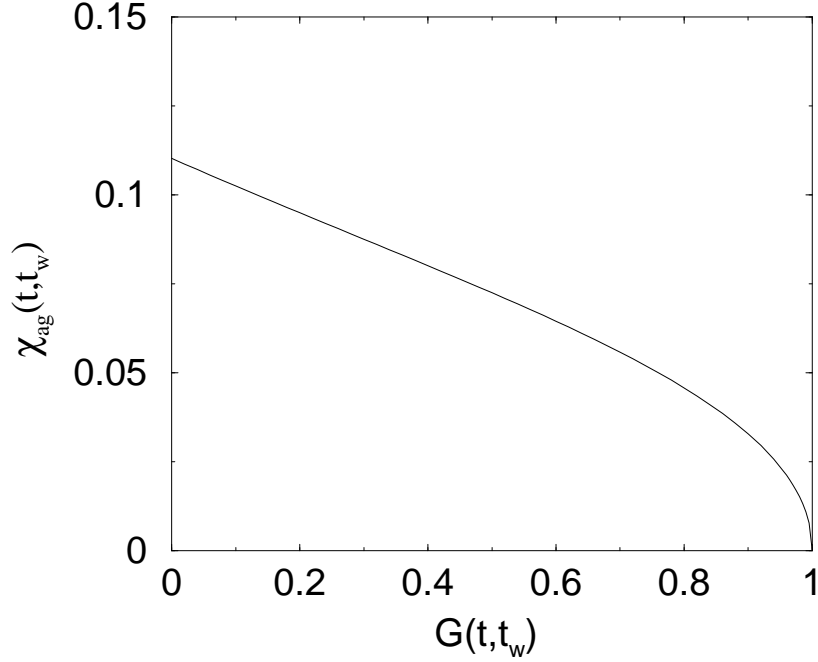,width=9cm,angle=-90}}
\caption{Parametric plot of $\chi_{ag}(t,t_w)$ vs. $G(t,t_w)$
for $d=2$}
\label{isinglike}
\end{figure}

For $d=2$ the power of $t_w$ in front of the integral  disappairs
from~(\ref{5.6}) and the leading contribution
for $t\gg \Lambda ^{-2}$ is given by
\be
\chi _{ag}(t,t_w)=\frac{1}{2\pi}\log \left (\frac{2}{1+\sqrt x}\right )
\label{a1.1}
\ee
showing that the aging contribution to the response does not vanish as 
$t_w\to \infty $. Furthermore, since $T_C=0$ for $d=2$, 
the phase-ordering process requires $T_F=0$ and from~(\ref{4.14})
follows
\be
G(t,t')=G_\sigma (t,t')=2M_0^2 \frac{\sqrt x}{1+x}.
\label{a2.1}
\ee
Eliminating $x$ between~(\ref{a1.1}) and~(\ref{a2.1}) we get
\be
\chi _{ag}(G)=\frac{1}{2\pi } \log \left \{
\frac{2}{1+\frac{M_0^2}{G}\left [ 1-\sqrt {1-\frac{G^2}{M_0^4}}\right ]}
\right \} 
\label{a2.2}
\ee
which gives a parametric plot (Fig.~\ref{isinglike}) qualitatively similar
to what one finds in the Ising model for $d=1$.

As it was recalled in the Introduction,
with a scalar order parameter the behavior of $\chi_{ag} (t,t_w)$
under variations of dimensionality becomes
trasparent by introducing the effective response $\chi_{eff}(t,t_w)$
associated to a single interface. 
The mechanism regulating the behavior of $\chi_{ag} (t,t_w)$
then is explained through the balance between the rate of loss
of interfaces as coarsening proceeds, and the rate of growth
of the single interface response given by~(\ref{i7}).
The dimensionality dependence of the exponent $\alpha $ in that case
is the outcome\cite{Corberi01} of the competition between the external field and
the curvature of interfaces in the drive of interface
motion.

\begin{figure}
\centerline{\psfig{figure=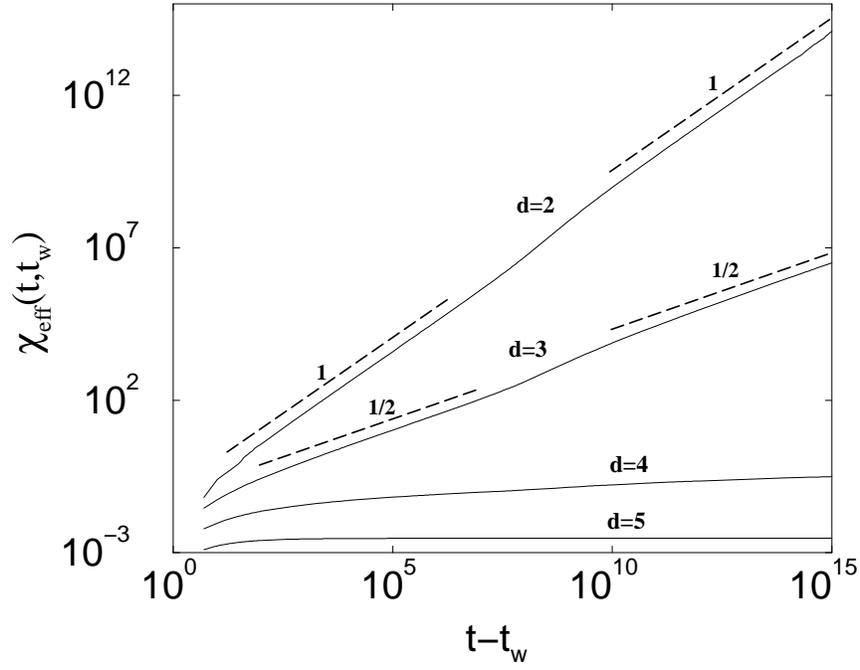,width=9cm,angle=-90}}
\caption{Plot of $\chi _{eff}(t,t_w)$ for $t_w=10^8$ and
$\Lambda =1$. The dashed lines are power laws with the 
corresponding exponent $\alpha $.}
\label{chieff}
\end{figure}

In the large $N$ model there are no localized defects and none of the
above concepts has direct physical meaning. Nonetheless, the notion
of defect density can be extended to the large $N$ case by looking
at the behavior of $I(t)$. Consider the quench to $T_F=0$ where
there are no thermal fluctuations. From~(\ref{3.14}), for large time,
$I(t)\sim -t^{-1}$ namely
\be
M_0^2-\frac{1}{N}\langle \vec \phi ^2 (\vec x,t)\rangle \sim t^{-1}.
\label{centodue}
\ee
If the system was in the ordered state, with the order parameter aligned 
everywhere, the left hand side ought to vanish. Therefore, the positive
difference~(\ref{centodue}) may be attributed to ``defects''
with density 
\be
\rho_D (t)\sim L^{-2}(t) \sim t^{-1}
\label{centotre}
\ee
which is what one obtains in general with a vector order 
parameter\cite{Bray94}.
We may then define the effective response function per defect by the
analogue of~(\ref{i6}) $\chi _{ag}(t,t_w)=\rho _D(t)\chi _{eff}(t,t_w)$
which yields
\be
\chi _{eff}(t,t_w)=t^{2-\frac{d}{2}}\int _x ^1 dy
\frac{y^{-\frac{d}{4}}-1}{\left ( 1-y+\frac{1}{\Lambda ^2 t} 
\right )^{\frac{d}{2}}}.
\label{centoquattro}
\ee
The behavior of this quantity can be computed analytically for short
and for large time separation obtaining in both cases, apart from
a change in the prefactor, a power law behavior as in~(\ref{i6})
\be
\chi _{eff}(t,t_w)\sim (t-t_w)^\alpha
\label{centocinque}
\ee
with
\be
\alpha = \left \{ \begin{array}{ll}
			0                    & \mbox{, for $d > 4$} \\  
			2-\frac{d}{2}         & \mbox{, for $d < 4$}
                   \end{array}                                        
               \right .
\label{centosei}
\ee
and
\be
\chi _{eff}(t,t_w)\sim \log \left [ \Lambda ^2 (t-t_w)\right ]
\label{centosette}
\ee
for $d=4$. The overall exact behavior of $\chi _{eff}(t,t_w)$
is depicted in Fig.\ref{chieff}, obtained by plotting~(\ref{centoquattro})
for different dimensionalities.

Therefore, the qualitative picture is the same obtained in the 
scalar case\cite{Corberi01}. 
There exist upper critical dimensionalities, $d_U=3$ for $N=1$ and
$d_U=4$ for $N=\infty$
(presumably for all $N>1$), above which $\chi _{eff}$ saturates
to a constant value within microscopic times. At $d_U$ $\chi _{eff}$
grows logarithmically, while below $d_U$ there is power law growth.
In addition there are lower critical dimensionalities, $d_L=1$ for
$N=1$ and $d_L=2$ for $N=\infty $ (or $N>1$), where the exponent $\alpha$ 
reaches exactly the value such that $\chi _{eff}(t,t_w)
\sim \rho _I^{-1}(t)$ in the scalar case and $\chi _{eff}(t,t_w)
\sim \rho _D^{-1}(t)$ in the vectorial case. Namely, at $d_L$ the 
growth of $\chi _{eff}$
makes up exactly for the loss of interfaces or defects, producing
the behavior illustrated in Fig.~\ref{isinglike}.

\section{Concluding remarks} \label{sec6}

In this paper we have shown that the basic features of the slow relaxation
phenomenology arising in phase ordering processes: separation of 
order parameter, autocorrelation function and linear response function
into fast and slow components, can be obtained analytically and
exactly in the large $N$ model. The behavior of $\chi _{ag}(t,t_w)$
is of particular interest since it displays the same qualitative pattern
of behavior under variation of dimensionality observed in the
Ising case. This is a strong indication that this might be a generic
feature of the slow relaxation in phase ordering processes. In this
respect it might be worth to undertake a numerical study of
$\chi _{ag}(t,t_w)$ in systems with vector order parameter and finite
$N$ at different dimensionalities.

A comment should be made about the connection between static and dynamic
properties. One of the most interesting recent developments in the
study of the off equilibrium deviation from FDT has been the 
derivation\cite{Franz98}
of a link between the response function and the structure of
the equilibrium state. Assuming that
$\lim_{t_w \rightarrow \infty} \chi (t,t_w) = \chi (G(t,t_w))$,
i.e. that in the large time regime $\chi (t,t_w)$ depends on
time only through the autocorrelation function, this connection
takes the form
\be
P(q) = \left . -T_F \frac{d^2 \chi (G)}{d G^2} \right ) _{G=q}
\label{c1}
\ee
where $P(q)$ is the probability that in the equilibrium state
the overlap $\frac{1}{NV} \int _V d\vec x \vec \phi (\vec x)
\cdot \vec \phi ' (\vec x)$ between two different configurations
$[\vec \phi (\vec x)]$ and $[\vec \phi ' (\vec x)]$ takes the value $q$.
For Ising systems~(\ref{c1}) holds for $d>1$. In fact, for $d>1$ the
$t_w \rightarrow \infty$ limit of $\chi (t,t_w)$ is 
found\cite{Barrat98,Berthier99,Corberi01} 
to have the form~(\ref{5.10}) which, applying~(\ref{c1}), yields
\be
P(q) = \delta (q - M^2)
\label{c2}
\ee
in agreement with an equilibrium state formed by the mixture of pure
states. For $d=1$, since $\chi _{ag} (t,t_w)$ does not disappear as
$t_w \rightarrow \infty$, the connection~(\ref{c1}) is no more 
valid\cite{Corberi01}. 
Then, in the large $N$ model we should expect~(\ref{c1}) to
fail at most for $d=2$, since that is the case where   
$\chi _{ag} (t,t_w)$ does not asymptotically vanish. However, it
is not so straightforward that~(\ref{c1}) should hold for
$d>2$. In fact, as observed above, the behavior of
$\chi (t,t_w)$ in the limit $t_w \rightarrow \infty$ 
for $d>2$ is indistinguishable
from what one obtains in the Ising case.
Namely, one finds the form~(\ref{5.10})  of $\chi(G)$ 
which yields~(\ref{c2}) for
$P(q)$ and this is what one expects when there is a mixture of ordered pure
states. The problem is, as explained in Section 2, that the structure of
the low temperature state in the large $N$ model is quite different
from a mixture of pure states. This puzzling
feature of the large $N$ model will be investigated in a separate paper.

Acknowledgments - This work has been partially supported
from the European TMR Network-Fractals c.n. FMRXCT980183 and 
from MURST through PRIN-2000.

\section{Appendix I}

Solving~(\ref{2.11}) in the large volume limit one finds:

\noindent for $0 <\frac{T_F - T_C}{T_C} \ll 1$

\be
(\xi \Lambda)^{-2}=\left \{ \begin{array}{ll}
                              \frac{\Lambda^2}{2}\left [\frac{M_0^2}{2T_F} 
                              \frac{(8\pi )^{d/2}}{\Gamma(1-d/2)}           
                              \right ]^{-\frac{2}{2-d}}              
                              &     \mbox{, for $ d<2 $} \\
                              e^{-\frac{4 \pi M_0^2}{T_F}}                                                     &     \mbox{, for $ d=2$ } \\
                              \left [\frac{T_F-T_C}{T_C} 
                              \frac{1}{\Gamma(2-d/2)}           
                              \right ]^{\frac{2}{d-2}}         
                              &     \mbox{, for $2<d<4$} \\
	                      \frac{T_F-T_C}{T_C}
                              \left ( \frac{\Lambda^2}{M_0^2g}+
                              \frac{T_F}{T_C}\frac{2}{d-4} \right )^{-1}	                                     &     \mbox{, for $d>4$}
                            \end{array}
                     \right .
\label{aI.1}
\ee

and

\be
   (\xi \Lambda)^{-2}\log \left ( (\xi \Lambda)^{-2} \right ) =
    - \left (\frac{T_F-T_C}{T_F} \right ) \quad 
    \mbox{, for $d=4$}.
\label{aI.2}
\ee

\noindent For $T_F=T_C$
\be
\xi =\left \{ \begin{array}{ll}
                              \frac{M_0^2}{T_C}\left (\frac{d-2}{d-4}
                              -\frac{\Lambda^2}{r} \right ) 
                              V^{\frac{1}{4}}               
                              &     \mbox{, for $ d>4 $} \\
                              \left [\Gamma(\frac{4-d}{2}) T_C^{\frac{d}{2}-1}
                              M_0^2 \Lambda^{2-d} \right ]^{\frac{1}{d}}
                              V^{\frac {1}{d}}
                              &     \mbox{, for $ d<4$ } 
                            \end{array}
                     \right .
\label{aI.3}
\ee
and
\be
\xi [ 2\log (\xi \Lambda)]^{-\frac{1}{4}} =
\left (\frac{M_0^2}{\Lambda^2 T_C} V \right )^{\frac{1}{4}}
\label{aI.4}
\ee
for $d=4$.

\section{Appendix II}

The prefactors in~(\ref{3.14}) are given by

\be
A_a=\left \{ \begin{array}{ll}
    \Delta \left [\frac{M_0^2}{\Gamma(1-d/2)} 
    \left ( \frac{8\pi}{T_F} \right )^{d/2}
    \right ]^{-\frac{2}{2-d}}                      & \mbox{, for $ d<2 $} \\
    2 \pi \frac{\Delta}{T_F^2}M_0^2 
    e^{-4 \pi \frac{M_0^2}{T_F}}                     & \mbox{, for $d=2$ }  \\
    \left [\frac{M_0^2}{T_C}(T_F-T_C) \right ]^{\frac{4-d}{2-d}}   
    \left ( \frac{1}{2g}+\frac{\Delta M_0^2}{2T_C} \right )
    \left [\frac{2T_F \vert \Gamma(1-d/2) \vert}{(8 \pi )^{d/2}}
    \right ]^{\frac{2}{d-2}}
    \frac{1}{d-2}                                  & \mbox{, for $2<d<4$} \\
     -\frac{(8 \pi)^2}{2T_C} \left ( \frac{1}{2g} +
    \frac{\Delta M_0^2}{2T_C} \right ) 
    \frac{l}{\log(2\xi ^{-2})}                              & \mbox{, for $d=4$} \\
    \left (\frac{1}{2g}+\frac{\Delta M_0^2}{2T_c} \right )
    \left ( \frac{1}{2g}+ \frac{M_0^2}{\Lambda^2}\frac{T}{T_c}
    \frac{1}{d-4} \right )^{-1}	                   & \mbox{, for $d>4$}
                              \end{array}
       \right .
\label{aII.1}
\ee
\be
A_c=\left \{ \begin{array}{ll}
    \frac{\Delta}{M_0^2} \frac{1}{(8\pi )^{d/2}}      & \mbox{, for $ d<2 $} \\
    \frac{\Delta}{8 \pi M_0^2}                        & \mbox{, for $d=2$ }  \\
    \sin \left [\left (\frac{d}{2}-1 \right) \pi \right ]
    \left ( \frac{1}{g}+\frac{\Delta M_0^2}{T_C} \right )
    \frac{(8\pi)^{d/2-1}}{T_C}(d-2)                  & \mbox{, for $2<d<4$} \\
    \frac{1}{2T_C} \left ( \frac{1}{2g} + 
    \frac{\Delta M_0^2}{2T_C} \right )               & \mbox{, for $d=4$}  \\
    \left (\frac{1}{2g}+\frac{\Delta M_0^2}{2T_C} \right )
    \left ( \frac{1}{2g}+ \frac{M_0^2}{\Lambda^2}
    \frac{1}{d-4} \right )^{-1}	                  &   \mbox{, for $d>4$}
                              \end{array}
       \right .
\label{aII.2}
\ee
and
\be
A_b=\frac{T_F+g\Delta M_0^2}{(8\pi )^{\frac{d}{2}}g\left [ M_0^2
\left ( 1-\frac{T_F}{T_C}\right ) \right ]^2}.
\label{aII.3}
\ee

\end{document}